\documentclass[prl,secnumarabic,amssymb, nobibnotes, aps, prd,aps,twocolumn]{revtex4-1}
\usepackage{graphicx , epsfig,color}
\usepackage{amsmath}
\usepackage{latexsym}
\usepackage{amstext}
\usepackage{array}
\usepackage{multirow}
\usepackage{changebar}
\usepackage{subfigure}
\usepackage{stackrel}

\newcommand{\ds}{\displaystyle}
\newcommand{\beq}{\begin{eqnarray}}
\newcommand{\eeq}{\end{eqnarray}}
\newcommand{\beqq}{\begin{eqnarray*}}
\newcommand{\eeqq}{\end{eqnarray*}}
\newcommand{\p}{\partial}

\newcommand{\eps}{\varepsilon}

\newcommand{\x}{\mbox{\boldmath$x$}}

\begin{document}

\title{Electrostatics of non-neutral biological microdomains}
\author{J. Cartailler$^{1}$,  Z. Schuss$^{2}$  and D. Holcman$^{1 3}$}
\affiliation{  $^1$ Ecole Normale Sup\'erieure, 46 rue d'Ulm 75005 Paris, France. $^2$ Department of Mathematics, Tel-Aviv University, Tel-Aviv 69978, Israel and $^3$ Mathematical Institute, University of Oxford, Oxford OX2 6GG, United Kingdom. }

\begin{abstract}
Voltage and charge distributions in cellular microdomains regulate communications, excitability, and signal transduction. We report here new electrical laws in a cell, which follow from a nonlinear electro-diffusion model. These newly discovered relations derive from the geometrical cell-membrane properties, such as membrane curvature, volume, and surface area. These
electro-diffusion laws can now be used to predict and interpret voltage distribution in cellular
microdomains.
\end{abstract}
\maketitle

Electro-diffusion in cellular microdmains remains difficult to study due to the lack of specific sensors and the theoretical hurdle of understanding charged particle in shaped geometrical domains. The diffusion of charged particles is largely influenced by the interaction of diffusing ions with the electrical field generated by all charges in the solution and possibly with external field.\\
The dielectric membrane of a charged biological cell also affects the electric field, because it creates image charges. So far, only a few electro-diffusion systems are well understood. For example, although the electrical battery was invented more than 200 years ago, designing optimal configurations is still a challenge. On the other extreme, ionic flux and gating of voltage-channels \cite{Bezanilla} is now well explained by the modern Poisson-Nernst-Planck (PNP) theory of electro-diffusion, because at the nanometer scale, cylindrical symmetry of a channel model reduces computations to a one-dimension model for the electric field and charge densities in the channel pore \cite{Eisenberg1,Eisenberg2,Roux1,EKS}. However, cellular microdomains involve two- and three-dimensional neuronal geometry \cite{Peskin,Savtchenko,Savtchenko1,Savtchenko2,Sejnowski}, which makes the analysis of the PNP equations much more complicated than in the cylindrical geometry of a channel pore.\\
We report here recent results about the distribution of charges and field, obtained from the analysis
of the nonlinear PNP model of electro-diffusion in various geometries of microdomains in the absence of electro-neutrality. In our model the entire boundary is impermeable to particles (ions) and the electric field satisfies the compatibility condition of Poisson's equation. Phenomenological descriptions of the electro-diffusion, such as cable equations or the reduced electrical-engineering approximation by resistance, capacitance, and even electronic devices, are insufficient for the description of non-cylindrical geometry \cite{HY2015}, because they assume simple one-dimensional or reduced geometry.\\
\begin{figure*}[http!]
	\center
	\includegraphics[scale=0.22]{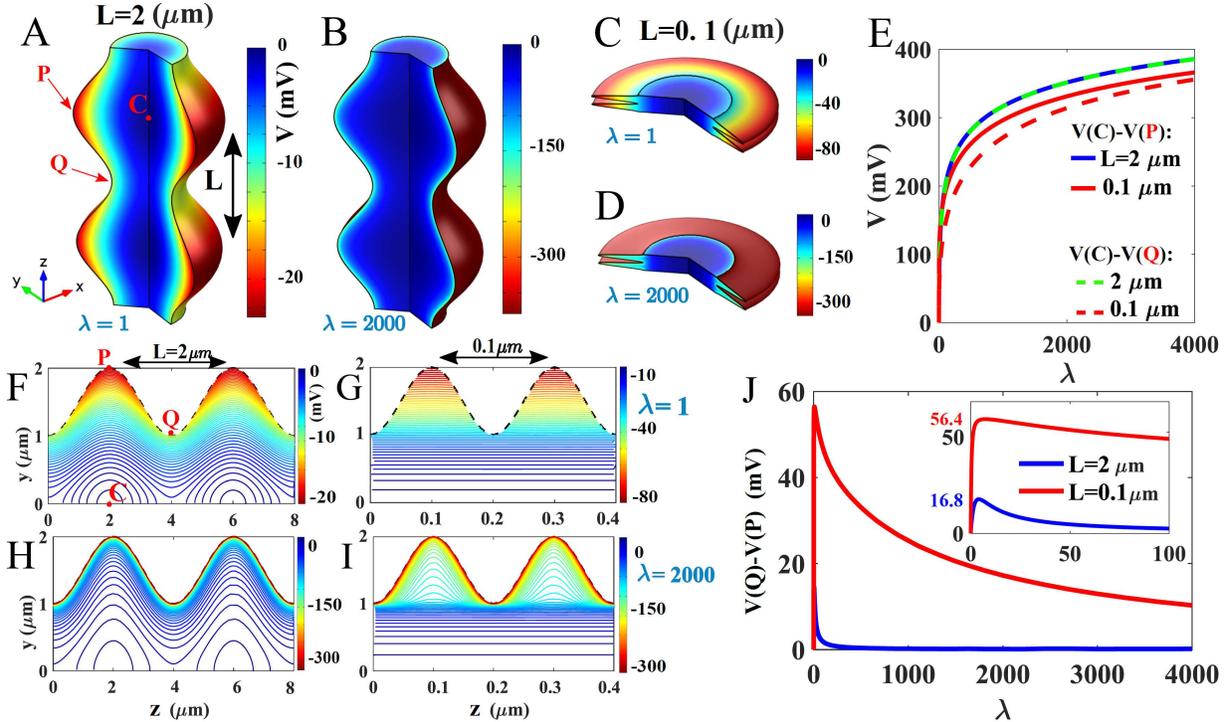}
	\caption{ { {\small Numerical evaluation of voltage distribution in a corrugated}
cylinder.  	{\bf A-B-C-D.} Voltage distribution computed for $\lambda=1$ ({\bf A-C}) and $\lambda=2000$ ({\bf B-D}). The boundary of the cylinder around the symmetry $z-$axis is defined by the curve $\gamma(z)=1+0.5\sin\left( 2\pi z/L \right)$, where $L$ is a parameter ($L=2\mu m$ {\bf A-B.}, and $L=0.1\mu m$ {\bf C-D}). {\bf E.} Voltage differences $V(C)-V(Q)$ (dashed line) and $V(P)-V(C)$ (solid line) vs. $\lambda$, computed for $L=0.1 \mu m$ (red) and $L=2\mu m$ (blue, green), where $P$ (resp.  $Q$) is the maximum (resp. minimum) of the curve $\gamma(z)$ and $C$ is defined by  $V(C)=\min(V)$.{\bf F-G-H-I.} Isopotential lines in the $yOz$ plane, computed for various $(\lambda,L)$: {\bf F.} $(1,2)$, {\bf G.} $(1,0.1)$, {\bf H.} $(2000,2)$, and {\bf I.} $(2000,0.1)$.
{\bf J.} Voltage difference $V(P)-V(Q)$ versus $\lambda$, computed for $L=2 \mu m$ (blue) and $L=0.1 \mu m$ (red). The inset in panel {\bf J} is a magnification of the small $\lambda$ region.}}		\label{Bandc}
\end{figure*}
{\bf\noindent Electrostatic theory with no electro-neutrality}\\
In the absence of electro-neutrality and with charge distributed in a bounded domain surrounded by a dielectric membrane, the PNP model for total charge $Q=zeN$, where $e$ is the electronic charge,  and the charge density  $\rho(\x,t)$, is given by \cite{Cartailler} for $\x\in\Omega,$
{\small
\beq
D\left[\Delta \rho(\x,t) +\frac{ze}{kT} \nabla \left(\rho(\x,t) \nabla V(\x,t)\right)\right]&=&\,
\frac{\p\rho(\x,t)}{\p t}\hspace{0.5em}\mbox{,}\ \label{NPE}\\
D\left[\frac{\p\rho(\x,t)}{\p n}+\frac{ze}{kT}\rho(\x,t)\frac{\p V(\x,t)}{\p n}\right]&=&\,0\hspace{0.5em}\mbox{,}\ \x\in\p\Omega \label{noflux}\\
\rho(\x,0)&=&\,q(\x)\hspace{0.5em}\mbox{,}\ \x\in\Omega,\label{IC}
\eeq }
where $V(\x,t)$ is the electric potential in $\Omega$. It is the solution of
the Poisson equation
\begin{align}
\label{poisson} \Delta V(\x,t) =
-\frac{ze\rho(\x,t)}{\eps\eps_0}\hspace{0.5em}\mbox{for}\ \x\in\Omega
\end{align}
with the boundary condition
\begin{align}\label{Boundary_Phi}
\frac{\p V(\x,t)}{\p
	n}=-\sigma(\x,t)\hspace{0.5em}\mbox{for}\ \x\in{\p\Omega}, 
\end{align}
where $\sigma(\x,t)$ is the surface charge density on the boundary $\p\Omega$. In the steady state,
\begin{align}
\sigma(\x,t)=\frac{Q}{\eps\eps_0 |\p \Omega|}.
\end{align}
Then \eqref{NPE} gives the density
{\small
\begin{align}
\rho(\x)=N\frac{\exp\left\{-\ds\frac{zeV(\x)}{kT}\right\}}
{\ds\int_\Omega\exp\left\{-\ds\frac{zeV(\x)}{kT}\right\}\,d\x},\label{N}
\end{align}}
hence \eqref{poisson} gives
\begin{align}
\Delta V(\x)=-\frac{zeN\exp\left\{-\ds\frac{zeV(\x)}{kT}\right\}}{\eps\eps_0{\ds\int_\Omega
		\exp\left\{-\ds\frac{zeV(\x)}{kT}\right\}\,d\x}},\label{Deltaphi}
\end{align}
and \eqref{Boundary_Phi} gives the boundary condition
{\small
\beq \label{compatibility}
\frac{\p V(\x)}{\p n} &=&-{ \frac{ \ds ze N}{\eps\eps_0|\p \Omega|}}\hspace{0.5em} \mbox{for}\ \x\in\p \Omega,
\eeq }
which is the compatibility condition, obtained by integrating Poisson's equation \eqref{poisson} over $\Omega$. \\
The concept of charge screening, which makes the induced field decay exponentially fast away from
a charge \cite{Debye1923}, does not apply and long-range correlations lead to a gradient of charges, as is the case, for example, in a ball without inward directed current. By solving \eqref{Deltaphi} numerically and asymptotically,  a new capacitance law was derived for an electrolytic solution in a ball \cite{Cartailler}, where the difference of potentials between the center and the surface, $V(C)-V(S)$, increases with the total number of charges, first linearly
and then logarithmically
\beq
V(C)-V(S) \approx {-2\frac{kT}{ze}\log\frac{(ze)^2N}{\eps\eps_0 kT}.}
\eeq
The effect of the geometry on the voltage and the charge distribution for other cell shapes is described below.\\
{\noindent \bf Local curvature at a cell boundary influences the field and charge distribution}\\
Axons and dendrites are not perfect cylinders and the curvature of their surfaces
has many local maxima \cite{Harris}. It turns out that this local curvature can influence the local voltage significantly, as shown in numerical solutions of the PNP equations (Fig. \ref{Bandc}), which reveal that regions of high curvature correspond to local charge accumulation. This effect should  be  sufficient to influence the voltage by creating a measurable local voltage increase of the order of a few millivolts. \\
The voltage can vary inside the cylinder and also along {curved surfaces} (Fig.
\ref{Bandc}A-B) and can depend on changes in curvature and in the total number of
charges $N$ (Fig. \ref{Bandc}C-F). Another interesting property of the curvature is that it creates a narrow boundary layer in the voltage (Fig. \ref{Bandc}D-F).\\
Note that the previous computations assume one specie only and in practice, due to the presence of other ions, $N$ or $\lambda= \frac{(ze)^2N}{\eps\eps_0 kT}$ should only represent the excess of positive charges. Thus in practice, we expect $\lambda\approx 10-50$, leading to voltage fluctuations of few mV. Consequently, the resting electric potential drop at cross-membrane and the resting voltage of voltage-gated channels along a dendrite may differ at different points, depending on curvature. This may affect the propagation and genesis of local depolarization or back propagation action potential in dendrites of neuronal cells \cite{Sackmann-BP1997}. \\
{\noindent \bf A cusp-shaped funnel can influence charge distribution}\\
\begin{figure*}[http!]
	\center
	\includegraphics[scale=0.18]{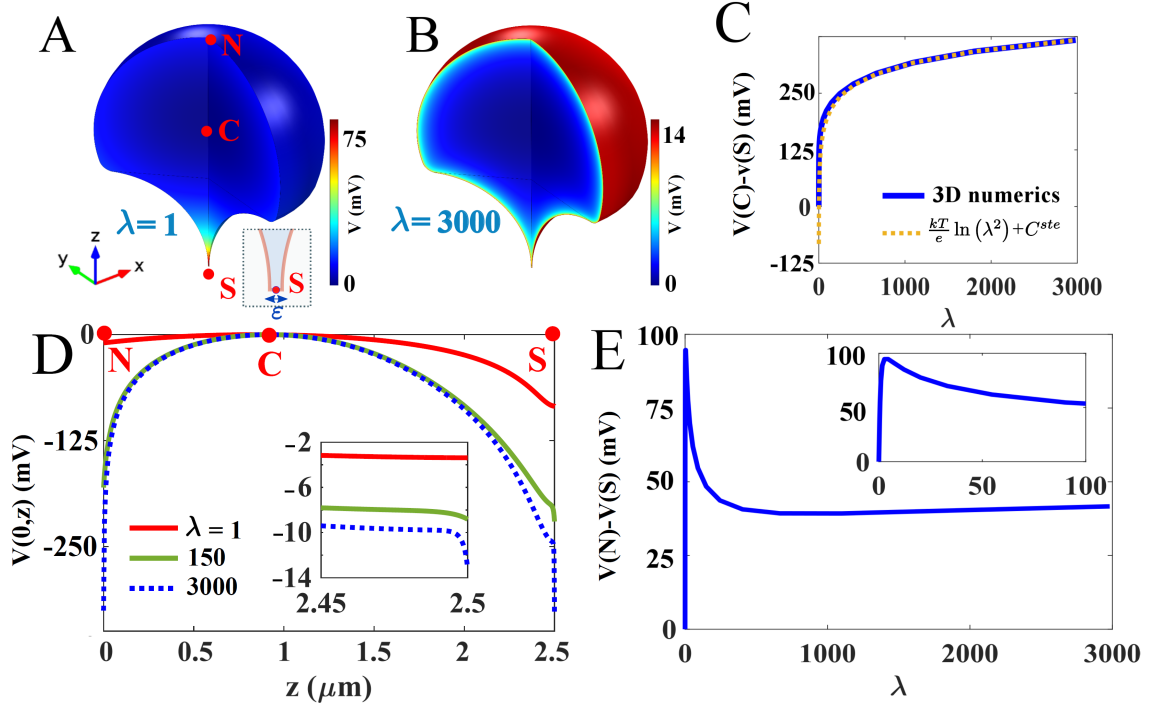}
	\caption{ {\small{Distribution of charge and voltage in a domain with a narrow funnel.} {\bf A-B.} Voltage distribution obtained for $\lambda=1$ ({\bf A}) and $\lambda=3000$ ({\bf B} ($\eps=0.01 \mu m$). {\bf C.} Difference of voltage between $V(C)-V(S)$ versus $\lambda$ (blue) compared to the logarithmic function $\frac{k T}{e}\ln{\lambda^2} +C^{ste}$, where $S$, $N$ and $C$ are the south, north pole and the center of mass respectively.
    {\bf D.} Voltage distribution evaluated along the $z-$axis, for $\lambda=1$ (red), $\lambda=150$ (green), and $\lambda=3\cdot10^3$ (dashed blue). The inset in panel {\bf D.} represents a magnification in the region of the pole $S$. {\bf E.} Voltage difference $V(N)-V(S)$ between the two poles versus $\lambda$. The inset in panel {\bf E} is a magnification in the small $\lambda$ region.}}
		\label{Band}
\end{figure*}
The connection between different cellular compartments is usually possible through narrow passages  that form cusp-shaped funnels, which have negative curvature (see the example described below) (Fig. \ref{Band}A).  We observe that the cusp-shaped funnel prevents the entrance of charges, at least when their number does not exceed a given threshold (Fig. \ref{Band}B-F). Therefore, there is a difference of steady state potential drop between the end of the funnel and the rest of the domain. \\
For a domain $\Omega$ with a cusp-shaped funnel $F$ formed by two bounding circles $A$ and $B$ of radii R (see Fig. \ref{Band}A), the opening of the funnel is $\eps\ll1$. The volume of the domain surface is $|\p \Omega|$. For a non-charged funnel the boundary condition \ref{Boundary_Phi} on the boundary of the funnel becomes  $\frac{\p V(\x,t)}{\p n}=0 \hbox{ for } \x \in F$. The difference of potential between the center of the two-dimensional domain and the end of the cusp is obtained by solving \ref{Deltaphi} using the conformal mapping method for the Laplace equation \cite{HS2015}, leading to:  for $\lambda\gg1,$
{\small
\beq\label{diffpp3dim}
V(S)-V(C)= \frac{kT}{ze}\log\left(\frac{2^9 R_c^{3/2} |\p   \Omega|\sqrt{ \eps}}{\pi^6 \lambda^2}\right) +O(1),
\eeq
}
where $R_c$ is the radius of curvature at the cusp. This difference of potential should be compared with the one between the north pole \cite{Cartailler} and the center given by
\beq\label{diffpp2dr1}
V(N)-V(C)=\frac{kT}{ze}\log \left(\frac{8\pi}{8\pi+\lambda_D}\right)^2.
\eeq
where  $\lambda_D= 2\pi\lambda R_l /|\p \Omega|$ ($R_l$ is the radius of the external domain $\Omega$). We conclude in the limit of $\lambda\gg1$ and $\tilde\eps\to0$ that the difference of potential between the end of cusp $S$ and the north pole $N$ in the domain is obtained by adding \eqref{diffpp2dr1} and \eqref{diffpp3dim} and we get
\beq\label{diffpp3dimpf}
V(S)-V(N)&=&\frac{kT}{ze}\ln\left(\frac{2^9 {R_c^{3/2}|\p \tilde \Omega| \sqrt{\tilde \eps}}}{\pi^6 }\right)\\
&+&2\frac{kT}{ze}\ln\left(\frac{R_l}{4|\p \tilde \Omega|}\right)+O\left(\frac1{\lambda}\right). \nonumber
\eeq
When the funnel is charged, the boundary condition inside the funnel is given by \ref{compatibility}, in that case, the difference of potential inside the cusp is
\beq
V(S)-V(N)=2\frac{kT}{ze}\ln\left( \frac{\pi c^2R_l }{16R_c}\right),
\eeq
where c is a constant that only depends on the geometry and $R_c$ the radius of curvature at the cusp.  At this stage, we conclude that the curvature at the cusp could be an interesting free parameter in the design of optimized nano-pipette to regulate the molecular and ionic fluxes. These pipettes were recently designed to record voltage in dendritic spines \cite{Krisna}.\\
\begin{figure*}[http!]
	\center
	\includegraphics[scale=0.3]{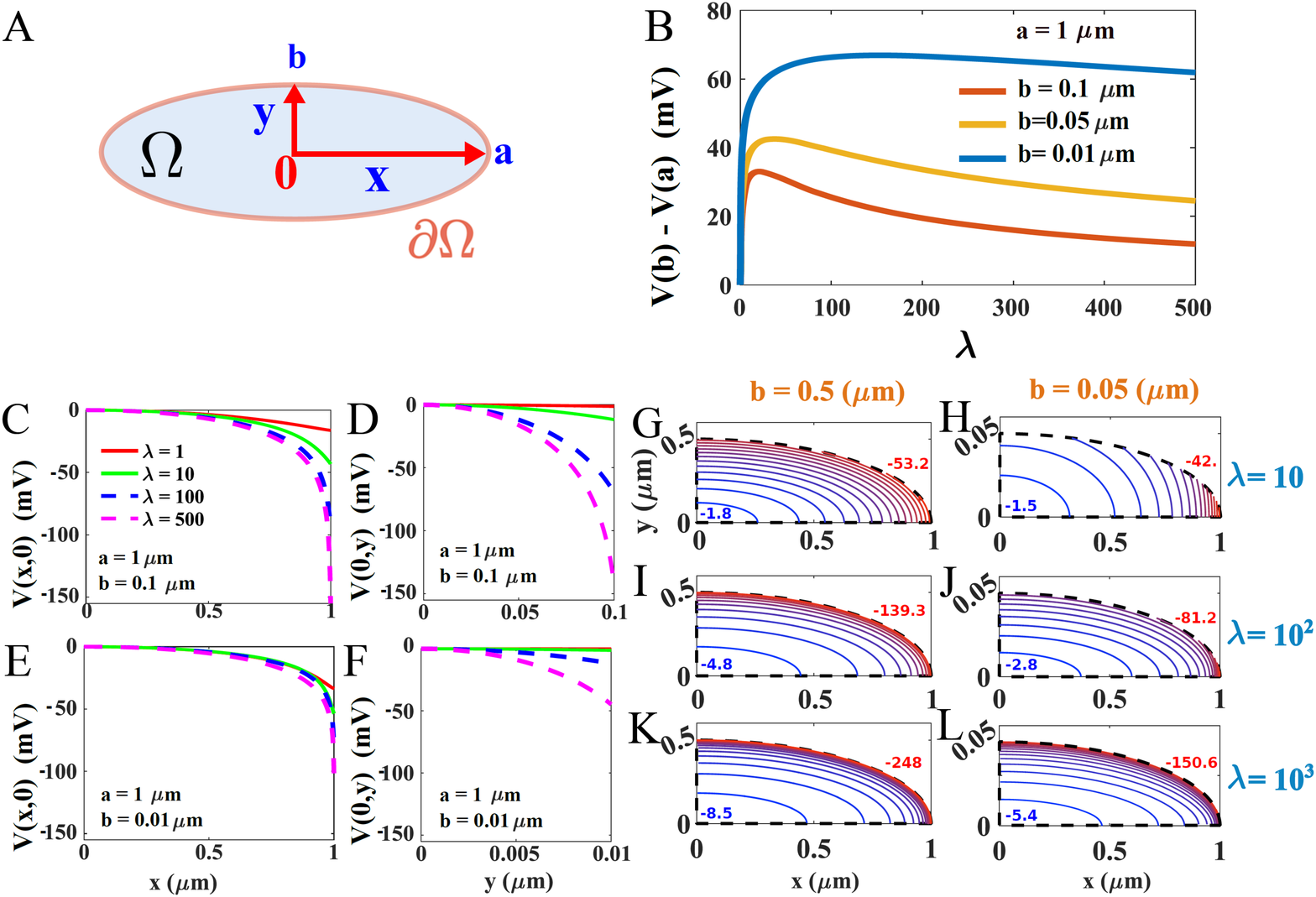}
	\caption{{\small {\bf Voltage distribution computed numerically in a narrow ellipse.} {\bf A.} Representation of the elliptic two-dimensional domain $\Omega$, where $a$ and $b$ are the major and the minor axis respectively. {\bf B.} Voltage difference $V(b)-V(a)$ versus $\lambda$, obtained for $a=1$ and $b=0.1$ (red), $b=0.05$ (yellow) and $b=0.01$ (blue). {\bf C-E} Voltage distribution computed along the $x-$axis for different values of $b$ and $\lambda$: $\lambda=1$ (red), $\lambda=10$ (green), $\lambda=100$ (dashed blue) and $\lambda=500$ (dashed magenta).  {\bf D-F} Voltage distribution computed along the minor axis $Oy$, for $b=0.1$ (panel {\bf D.}) and $b=0.01$ (panel {\bf F.}). {\bf G-L.} Equipotential contours computed for $a=1$ and varying $(b,\lambda)$. The red values represent the minimal potential while the blue ones are the maximal potential.}}
\label{Fig_Ellipse}
\end{figure*}
{\noindent \bf Voltage in a flat ellipsoidal electrolyte} \\
Numerical solutions of the PNP equations \ref{Deltaphi} (Fig. \ref{Fig_Ellipse}A) show the effect of local charge accumulation in a flat ellipse in two dimensions, in particular there is a maximum of the difference of potential $V(A)-V(B)$ between the end point of the small and long ellipse axis (fig. \ref{Fig_Ellipse}B-C). The voltage along each axis is shown in fig. \ref{Fig_Ellipse}D-G.  The elliptic domain  can represent the cross section of an axon, and confirms the effect of curvature discussed in the previous sections. Specifically, that charges accumulate near the boundary of highest curvature. The equipotential contours are shown in Fig. \ref{Fig_Ellipse}G-L.\\

To conclude, local changes in geometry, non-electro-neutrality, and dielectric boundary affect charge distribution {in electro-diffusion of} electrolytes, as shown here and \cite{Nitzan,HY2015,Cartailler}.We presented here a model with only one specie, but in reality multiple positive and negative ions are mixed in solution. We applied the present electro-diffusion model to capture the consequence of an excess of positive ions. Consequently, the large value for the voltage and voltage differences that we obtained for the large charge limit $\lambda \rightarrow \infty$ are probably attenuated in a mixed ionic solution, when the electro-neutrality remains broken.\\
The present results should be of significant consequences for neurons, sensory and glial cells, and many more. Indeed, local curvature is associated with gradient of charge density that can affect the electrical properties of micro-compartments \cite{YusteBook,Araya1,Araya2}. In all cases, the density of charge accumulates near boundary points of locally maximal curvature. These results can further be used to design nano-devices such as pipettes and to better understand voltage changes inside dendrites and axons. Future analysis should reveal charge distribution during transient current.\\
Acknowledgement: DH research was supported by a Marie-Curie Award grant.
\bibliographystyle{plain}

\begin{thebibliography}{99}

\bibitem{Bezanilla} Bezanilla, F. How membrane proteins sense voltage. N\textit{at Rev Mol Cell Biol.} \textbf{9}, 323-32 (2008).

\bibitem{Eisenberg1}Eisenberg, R.S,  (1998).``Ionic channels in biological membranes. Electrostatic analsis of a natural nanotube."   \textit{Contemp. Phys.}, {\bf39} (6), pp.447--466.

\bibitem{Eisenberg2}Eisenberg, R.S., (1998). ``From structure to function in open ionic channels."  \textit{J. Membrane Biol.}, {\bf171}, pp.1--24.

\bibitem{Roux1} Horn R., Roux B., Aqvist J. Permeation redux: thermodynamics and kinetics of ion movement through potassium channels. \textit{Biophys J.}, {\bf106}(9), 1859-63 (2014)

\bibitem{EKS}Eisenberg, R.S., M.M. Klosek, and Z. Schuss, (1995).
``Diffusion as a chemical reaction: Stochastic trajectories between fixed
concentrations." \textit{J. Chem. Phys.}, {\bf102} (4), pp.1767--1780.

\bibitem{Peskin} Lee P, Sobie EA, Peskin CS, Computer simulation of voltage sensitive calcium ion channels in a dendritic spine, J. Theor Biol. 2013;338:87-93.

\bibitem{Sejnowski} Lopreore CL, Bartol TM, Coggan JS, Keller DX, Sosinsky GE, Ellisman MH, Sejnowski TJ.
Computational modeling of three-dimensional electrodiffusion in biological systems: application to the node of Ranvier. Biophys J. 2008 Sep 15;95(6):2624-35.

\bibitem{Savtchenko}Savtchenko LP, Kulahin N, Korogod SM, Rusakov DA, Electric fields of synaptic currents could influence diffusion of charged neurotransmitter molecules. Synapse. 2004,  15;51(4):270-8.

\bibitem{Savtchenko1} Sylantyev S, Savtchenko LP, Niu YP, Ivanov AI, Jensen TP, Kullmann DM, Xiao MY, Rusakov DA, Electric fields due to synaptic currents sharpen excitatory transmission.Science. 2008 28;319(5871):1845-9.

\bibitem{Savtchenko2} Sylantyev, S., Savtchenko, L.P., Ermolyuk, Y., Michaluk, P., and Rusakov, D.A. (2013). Spike-driven glutamate electrodiffusion triggers synaptic potentiation via a homer-dependent mGluR-NMDAR link. Neuron 77, 528-541.

\bibitem{HY2015}Holcman, D. and R. Yuste (2015).``The new nanophysiology: regulation of ionic flow in neuronal subcompartments," \textit{Nature Reviews Neuroscience} \textbf{16}, pp.685--692.

\bibitem{Cartailler} J Cartailler, Z Schuss, D Holcman,  Analysis of the Poisson–Nernst–Planck equation in a ball for modeling the Voltage–Current relation in neurobiological microdomains, (2016) Physica D: Nonlinear Phenomena 339, 39-48 (2017).

\bibitem{Debye1923} P. Debye and E. H\"uckel, Zur Theorie der Elektrolyte. I. Gefrierpunktserniedrigung und verwandte Erscheinungen, Physikalische Zeitschrift, Vol. 24, No. 9, 1923, p. 185-206

\bibitem{Harris}Bourne JN, Harris KM., Balancing structure and function at hippocampal dendritic spines. Annu Rev Neurosci. 2008;31:47-67.

\bibitem{Sackmann-BP1997}Markram H, Lubke J, Frotscher M, Sakmann B., Regulation of synaptic efficacy by coincidence of postsynaptic APs and EPSPs. Science. 1997;275(5297):213-5.

\bibitem{HS2015}Holcman, D. and Z. Schuss (2015), \textit{Stochastic Narrow Escape in Molecular an Cellular Biology, Analysis and Applications}, Springer Verlag, NY 2015.

\bibitem{Krisna}Jayant K, Hirtz JJ, Plante IJ, Tsai DM, De Boer WD, Semonche A, Peterka DS, Owen JS, Sahin O, Shepard KL, Yuste R., Targeted intracellular voltage recordings from dendritic spines using quantum-dot-coated nanopipettes. Nat Nanotechnol. 2016

\bibitem{Nitzan}Mamonov, A., R. Coalson, A., Nitzan, M., Kurnikova, (2003). "The role of the dielectric barrier in narrow biological channels: a novel composite approach to modeling single channel currents", \textit{Biophys. J.} {\bf84}, pp.3646--3661.

\bibitem{YusteBook}Yuste, R. (2010). \textit{Dendritic Spines}, The MIT Press, Cambridge, MA.

\bibitem{Araya1} Araya, R., Jiang, J., Eisenthal, K.B., Yuste, R. The spine neck filters membrane potentials.\textit{ Proc. Natl. Acad. Sci. USA} \textbf{103}, 17961-17966 (2006).

\bibitem{Araya2} Araya R, Vogels TP, Yuste R. Activity-dependent dendritic spine neck changes are correlated with synaptic strength.\textit{ Proc. Natl. Acad. Sci. USA} \textbf{111}, 2895-904. (2014).


\end{thebibliography}

\end{document}